\documentclass[10pt,a4paper]{article}

\usepackage[utf8]{inputenc}
\usepackage[english]{babel}

\begin{document}
\large

\newpage
\begin{center}
\LARGE{\bf On a Mass-Charge Structure of Gauge Invariance}
\end{center}
\vspace{0.1mm}
\begin{center}
{\bf Rasulkhozha S. Sharafiddinov}
\end{center}
\vspace{0.1mm}
\begin{center}
{\bf Institute of Nuclear Physics, Uzbekistan Academy of Sciences,
\\Tashkent, 100214 Ulugbek, Uzbekistan}
\end{center}
\vspace{0.1mm}

\begin{center}
{\bf Abstract}
\end{center}

The mathematical logic of a true nature of mirror symmetry expresses, in the case of the 
Dirac Lagrangian, the ideas of the left- and right-handed photons referring to long- and 
short-lived particles, respectively. Such a difference in lifetimes says about the photons 
of the different components having the unidentical masses, energies, and momenta. This requires 
the generalization of the classical Klein-Gordon equation to the case of all types of bosons 
with a nonzero spin. The latter together with a new Dirac equation admits the existence of the 
second type of the local gauge transformation responsible for origination in the Lagrangian of 
an interaction Newton component, which gives an inertial mass to all the interacting matter fields. The quantum mass operator and the mirror presentation of the classical Schr\"odinger equation 
suggest one more highly important equation. Findings show clearly that each of the quantum mass, energy, and momentum operators can individually act on the wave function. They constitute herewith 
the Euler-Lagrange equation at the level of the mass-charge structure of gauge invariance.

\vspace{0.6cm}
\noindent
{\bf 1. Introduction}
\vspace{0.3cm}

One of the most highlighted features of symmetry laws is their unity, which states that the left (right)-handed neutrino in the field of emission can be converted into the right (left)-handed one without change of his lepton flavor. These transitions indicate to the existence of the difference 
in masses, energies, and momenta of neutrinos of the different components. As was noted, however, by the author in [1] for the first time, such a possibility is realized owing to the spontaneous mirror symmetry violation. Thereby, it expresses the ideas about that the left-handed neutrino and the right-handed antineutrino are of long-lived leptons, and the right-handed neutrino and the 
left-handed antineutrino refer to short-lived fermions. 

This in turn allows to establish a new CP-odd Dirac equation, which involves that the mass, energy, and momentum come forward in nature of vector types of particles as the flavor symmetrical matrices
\begin{equation}
m_{s}={{m_{V} \, \, \, \, 0}\choose{\ 0 \, \, \, \, \ m_{V}}}, \, \, \, \,
E_{s}={{E_{V} \, \, \, \, 0}\choose{\ 0 \, \, \, \, \ E_{V}}}, \, \, \, \,
{\bf p}_{s}={{{\bf p}_{V} \, \, \, \, 0}\choose{\ 0 \, \, \, \, \ {\bf p}_{V}}},
\label{1}
\end{equation}
\begin{equation}
m_{V}={{m_{L} \, \, \, \, 0}\choose{\ 0 \, \, \, \, \ m_{R}}}, \, \, \, \,
E_{V}={{E_{L} \, \, \, \, 0}\choose{\ 0 \, \, \, \, \ E_{R}}}, \, \, \, \,
{\bf p}_{V}={{{\bf p}_{L} \, \, \, \, 0}\choose{\ 0 \, \, \, \, \ {\bf p}_{R}}},
\label{2}
\end{equation}
where $V$ must be accepted as an index of a distinction.

We recognize that between the currents of C-even and C-odd charges there exists a range of  
paradoxical contradictions [2], which require their classification with respect to C-operation. 
It reflects the availability of the two types of particles and fields of vector $V_{l}$ and 
axial-vector $A_{l}$ currents of the different C-parity [3]. 

The mass, energy, and momentum of a C-odd neutrino are strictly axial-vector $(A)$ type [4]. In contrast to this, a C-even neutrino has the mass, energy, and momentum of a vector $(V)$ nature [1]. In other words, the matrices (\ref{1}) and (\ref{2}) refer to those neutrinos 
$(\nu_{l}^{V}=\nu_{e}^{V},$ $\nu_{\mu}^{V},$ $\nu_{\tau}^{V}, ...)$ and leptons 
$(l^{V}=e^{V},$ $\mu^{V},$ $\tau^{V}, ...)$ in which axial-vector C-odd properties are absent. 

From their point of view, well known [5] Dirac equation itself for vector types of particles with 
the spin $1/2$ and the four-component wave function $\psi_{s}(t_{s}, {\bf x}_{s})$ accepts 
a latent united form
\begin{equation}
i\frac{\partial}{\partial t_{s}}\psi_{s}=\hat H_{s}\psi_{s}.
\label{3}  
\end{equation} 
Here
\begin{equation}
\hat H_{s}=\alpha \cdot\hat {\bf p_{s}}+\beta m_{s},
\label{4}
\end{equation}
and $m_{s},$ $E_{s},$ and ${\bf p}_{s}$ describe in a mirror presentation [1] the quantum 
mass, energy, and momentum operators 
\begin{equation}
m_{s}=-i\frac{\partial}{\partial \tau_{s}}, \, \, \, \,  
E_{s}=i\frac{\partial}{\partial t_{s}}, \, \, \, \,  
{\bf p}_{s}=-i\frac{\partial}{\partial {\bf x}_{s}}.
\label{5}
\end{equation}

As in Eq. (\ref{1}), an index $s$ implies that the space-time coordinates $(t_{s}, {\bf x}_{s})$
and the lifetimes $\tau_{s}$ for the left $(s=L=-1)$- and right $(s=R=+1)$-handed particles 
are different.

In these circumstances any of earlier experiments [6] about sterile neutrinos, namely, about 
right-handed short-lived neutrinos testifies [1] in favor of a vector mirror Minkowski space-time. 

If we now use the Dirac matrices $\gamma^{\mu}=(\beta, \beta \alpha)$ on account of 
$\partial_{\mu}^{s}=\partial/\partial{\it x}^{\mu}_{s}=(\partial/\partial t_{s}, -\nabla_{s}),$ 
Eq. (\ref{3}) is reduced to the following form:
\begin{equation}
(i\gamma^{\mu}\partial_{\mu}^{s}-m_{s})\psi_{s}=0
\label{6}
\end{equation}
in which $\partial_{\mu}^{s}$ is predicted as 
\begin{equation}
\partial_{\mu}^{s}=
{{\partial_{\mu}^{V} \, \, \, \, 0}\choose{\ 0 \, \, \, \, \ \partial_{\mu}^{V}}}, \, \, \, \,
\partial_{\mu}^{V}=
{{\partial_{\mu}^{L} \, \, \, \, 0}\choose{\ 0 \, \, \, \, \ \partial_{\mu}^{R}}}.
\label{7}
\end{equation}

This does not imply of course that the mass, energy, and momentum of the neutrino of a vector 
current at the level as were united by the author [1] in a unified whole do not transform the 
left (right)-handed neutrino into a right (left)-handed one without violate of Lorentz symmetry. 
We encounter, thus, the fact [3] that regardless of whether or not an unbroken Lorentz invariance exists, the same neutrino may not be simultaneously both a left-handed fermion and a right-handed one. At the same time, each type of gauge boson constitutes a kind of physical current, which 
follows from the unified mirror principle.

Therefore, to understand the mathematical logic of a C-invariant nature of matter at the new 
fundamental dynamical level, one must establish a mirror picture of the united interactions and 
a role in their formation of a latent structure of gauge invariance including a unified theoretical 
description of the origination of a mass of vector types of particles and fields. 

For this purpose, we investigate in a given work the questions implied from the invariance
of the free Dirac Lagrangian
\begin{equation}
L_{free}^{D}=\overline{\psi}_{s}(i\gamma^{\mu}\partial_{\mu}^{s}-m_{s})\psi_{s}
\label{8}  
\end{equation} 
concerning the two types of the local vector gauge transformations.

\vspace{0.6cm}
\noindent
{\bf 2. An equation for bosons with a nonzero spin}
\vspace{0.3cm}

One of the local vector types of gauge transformations can appear in the charge structure dependence of gauge invariance and in the Coulomb $(C)$ limit is written in the form
\begin{equation}
\psi'_{s}=U_{s}^{C}\psi_{s}, \, \, \, \, U_{s}^{C}=e^{i\alpha_{s}(x_{s})}
\label{9}  
\end{equation} 
and that, consequently, at the locality of its self phase $\alpha_{s}(x_{s}),$ the Lagrangian (\ref{8}) becomes gauge noninvariant.

It is not excluded, however, that this broken symmetry may be restored if we introduce the photon Coulomb field $A_{\mu}^{s}(x_{s}),$ which must have the following gauge transformation: 
\begin{equation}
A_{\mu}^{s'}=A_{\mu}^{s}+\frac{i}{e_{s}}\partial_{\mu}^{s}\alpha_{s},
\label{10}  
\end{equation} 
where $e_{s}$ denote the Coulomb mirror interaction constants at the level of an electric charge 
of a C-invariant nature.

Supposing in Eq. (\ref{8}) that 
\begin{equation}
\partial_{\mu}^{s}=\partial_{\mu}^{s}-e_{s}A_{\mu}^{s},
\label{11}  
\end{equation} 
it is easy to observe the invariance of the Dirac Lagrangian
$$L^{D}=L_{free}^{D}+L_{int}^{D}=$$
\begin{equation}
=\overline{\psi}_{s}(i\gamma^{\mu}\partial_{\mu}^{s}-m_{s})\psi_{s}-
ie_{s}\overline{\psi}_{s}\gamma^{\mu}\psi_{s}A_{\mu}^{s}
\label{12}  
\end{equation} 
concerning the action of chosen gauge transformations (\ref{9}) and (\ref{10}) due to the 
interaction with the photon Coulomb field of CP-odd types of fermions. 

However, as is now well known, quantum operator $\partial_{\mu}^{s}$ is $4\times 4$ matrix,
and consequently, the field $A_{\mu}^{s}$ equalized with the Coulomb field of a vector photon $(\gamma^{V})$ is not of those gauge fields in which the mass is absent. It can therefore be expressed in the form
\begin{equation}
A_{\mu}^{s}=\pmatrix{A_{\mu}\cr B_{\mu}}, \, \, \, \,
A_{\mu}=\pmatrix{A_{\mu}^{L}\cr A_{\mu}^{R}}, \, \, \, \,
B_{\mu}=\pmatrix{B_{\mu}^{L}\cr B_{\mu}^{R}}.
\label{13}
\end{equation}

This new mirror presentation of the vector photon fields of the Coulomb nature corresponds in 
the Lagrangian (\ref{12}) to the fact [1] that in them
\begin{equation}
\psi_{s}=\pmatrix{\psi\cr \phi}, \, \, \, \,
\psi=\pmatrix{\psi_{L}\cr \psi_{R}}, \, \, \, \,
\phi=\pmatrix{\phi_{L}\cr \phi_{R}}.
\label{14}
\end{equation}

The compound structure of both types of fields (\ref{13}) and (\ref{14}) testifies that the Lagrangian $L_{int}^{D}$ in Eq. (\ref{12}) expresses the ideas of earlier experiments. These ideas
require the definition of neutrino helicity [7-9] in the spin direction dependence of those photons,
which at the nuclear $\beta$-decay suffer resonant scattering. Then it is possible, for example, to interpret the implications of each of the mentioned experiments [7-9] about neutrino helicity as the group of arguments in favor of that the left (right)-handed neutrinos due to the spontaneous mirror symmetry violation have no interaction with right (left)-handed photons. They possess with all the left (right)-handed gauge bosons the same interaction as the electrons of a left (right) helicity.

According to such a principle, the left- and right-handed vector photons refer to long-and 
short-lived particles, respectively. This difference in lifetimes of photons of a definite 
helicity can explain the spontaneous mirror symmetry violation, which comes forward in the 
universe of C-invariant types of particles and fields as a spontaneity criterion of gauge 
invariance violation [10]. Thereby, it says about that the vector photons of the different components have the unidentical masses, energies, and momenta. We can therefore generalize [11] the classical Klein [12]-Gordon [13] equation from the quantum electrodynamics of spinless particles [14] to the case of vector types of bosons with a nonzero spin. One can define the structure of the latter for the four-component wave function $\varphi_{s}(t_{s}, {\bf x}_{s})$ in a mirror presentation by the following manner:
\begin{equation}
(\partial_{\mu}^{s}\partial^{\mu}_{s}+m_{s}^{2})\varphi_{s}=0,
\label{15}
\end{equation}
where we must use one more highly important connection
\begin{equation}
\varphi_{s}=\pmatrix{\varphi\cr \chi}, \, \, \, \,
\varphi=\pmatrix{\varphi_{L}\cr \varphi_{R}}, \, \, \, \,
\chi=\pmatrix{\chi_{L}\cr \chi_{R}}.
\label{16}
\end{equation}

So it is seen that the free boson Lagrangian responsible for a new equation (\ref{15}) may 
be presented in the form 
\begin{equation}
L_{free}^{B}=
\frac{1}{2}\varphi_{s}^{*}(\partial_{\mu}^{s}\partial^{\mu}_{s}+m_{s}^{2})\varphi_{s}.
\label{17}  
\end{equation} 

Of course, this Lagrangian is noninvariant at his self local phase $\alpha_{s}(x_{s})$ of 
a vector gauge transformation such as 
\begin{equation}
\varphi'_{s}=U_{s}^{C}\varphi_{s}, \, \, \, \, U_{s}^{C}=e^{i\alpha_{s} (x_{s})}.
\label{18}  
\end{equation} 

For restoration of its broken symmetry, it is desirable to introduce again the field $A_{\mu}^{s}$ with the corresponding gauge transformation. 

Inserting Eq. (\ref{11}) in Eq. (\ref{17}), it can be easily verified that at the local gauge 
transformations (\ref{10}) and (\ref{18}), the Lagrangian
$$L^{B}=L_{free}^{B}+L_{int}^{B}=$$
$$=\frac{1}{2}\varphi_{s}^{*}(\partial_{\mu}^{s}\partial^{\mu}_{s}+m_{s}^{2})\varphi_{s}+$$
\begin{equation}
+\frac{1}{2}[e_{s}(\varphi_{s}^{*}\partial_{\mu}^{s}\varphi_{s}A^{\mu}_{s}-
\varphi_{s}^{*}\partial^{\mu}_{s}\varphi_{s}A_{\mu}^{s})-
e_{s}^{2}\varphi_{s}^{*}\varphi_{s}A_{\mu}^{s}A^{\mu}_{s}]
\label{19}  
\end{equation} 
remains invariant concerning their action. 

\vspace{0.6cm}
\noindent
{\bf 3. Unity of Coulomb and Newton types of matter fields}
\vspace{0.3cm}

If we now take into account that the presence of the right-handed particles and fields in 
Lagrangians (\ref{12}) and (\ref{19}) is by no means excluded naturally, then there arises a 
question of whether the zero value of any of the fermion or the photon masses is not strictly
nonverisimilar even at their interaction with the field of emission of the photon itself.

Here an important circumstance is that the Lagrangian $L_{int}^{D}$ in Eq. (\ref{12}) describes only 
the part of the Dirac interaction, which corresponds to the electric charges of the interacting objects. On the other hand, as was noted in [15] for the first time, any interaction between the fermion and the field of emission includes both a kind of Newton and a kind of Coulomb components. 
At the same time, the Dirac interaction itself must contain simultaneously each of both types of 
the structural parts. This becomes possible owing to a mass-charge duality [16], according to which, any of electric $E,$ weak $W,$ strong $S,$ and other types of charges testifies in favor of the availability of a kind of inertial mass. The particle mass and charge are united in rest mass $m_{s}^{U}$ and charge $e_{s}^{U}$ equal to all the mass and charge
\begin{equation}
m_{s}=m_{s}^{U}=m_{s}^{E}+m_{s}^{W}+m_{s}^{S}+...,
\label{20}
\end{equation}
\begin{equation}
e_{s}=e_{s}^{U}=e_{s}^{E}+e_{s}^{W}+e_{s}^{S}+....
\label{21}
\end{equation}

It is already clear from the foregoing that each free Lagrangian consists of Newton $N$ 
and Coulomb $C$ components, the invariance of any of which concerning a kind of local gauge transformation admits the appearance of the corresponding part of the same interaction. Such 
a correspondence principle expresses, in the case of $m_{s}=m_{s}^{E}$ and $e_{s}=e_{s}^{E},$ 
the invariance of each Lagrangian $L^{D}$ or $L^{B}$ at the action of one more another type of 
the local vector gauge transformation responsible for the existence in them of an interaction Newton
component. This second type of a vector transformation having in the limits of fermion $\psi_{s}$ and boson $\varphi_{s}$ fields the different local phase $\alpha_{s}(\tau_{s})$ can appear in the mass structure dependence of gauge invariance and behaves as follow:
\begin{equation}
\psi'_{s}=U_{s}^{N}\psi_{s}, \, \, \, \, U_{s}^{N}=e^{i\alpha_{s}(\tau_{s})},
\label{22}  
\end{equation} 
\begin{equation}
\varphi'_{s}=U_{s}^{N}\varphi_{s}, \, \, \, \, U_{s}^{N}=e^{i\alpha_{s}(\tau_{s})}.
\label{23}  
\end{equation} 

At first sight, such a conclusion does not correspond to reality at all. But we can add that this 
is not quite so. The point is that a Newton part with mass $m_{s}$ in any of Lagrangians (\ref{8}) and (\ref{17}) is general concerning the corresponding gauge transformation (\ref{9}) or (\ref{18})
and does not depend of whether it has a local or a global phase. In the same way, one can see that 
all conditions of gauge symmetry of a Coulomb component with operator $\partial_{\mu}^{s}$ hold 
in each of Lagrangians (\ref{8}) and (\ref{17}) regardless of whether the suggested second type 
of transformation is or not present in it.

Such a picture is not changed even at the choice of a particle mass $m_{s}=-i\partial_{\tau}^{s},$
because of which Eq. (\ref{6}) is reduced to another new equation
\begin{equation}
(\gamma^{\mu}\partial_{\mu}^{s}+\partial_{\tau}^{s})\psi_{s}=0
\label{24}
\end{equation}
implied from the free Dirac Lagrangian
\begin{equation}
L_{free}^{D}=i\overline{\psi}_{s}(\gamma^{\mu}\partial_{\mu}^{s}+\partial_{\tau}^{s})\psi_{s},
\label{25}  
\end{equation} 
where we must keep in mind that
\begin{equation}
\partial_{\tau}^{s}=
{{\partial_{\tau}^{V} \, \, \, \, 0}\choose{\ 0 \, \, \, \, \ \partial_{\tau}^{V}}}, \, \, \, \,
\partial_{\tau}^{V}=
{{\partial_{\tau}^{L} \, \, \, \, 0}\choose{\ 0 \, \, \, \, \ \partial_{\tau}^{R}}}.
\label{26}
\end{equation}

These connections would seem to say that either $\psi_{s}$ behaves as a function of the fermion 
lifetimes and space-time coordinates $\psi_{s}(t_{s}, {\bf x}_{s}, \tau_{s})$ or the hypothesis 
about the quantum operator presentation of their mass is not valid. On the other hand, as follows from considerations of mirror symmetry, spin properties of a particle depend not only on its mass, energy, and momentum but also on the nature of space [1], where it is characterized by space-time coordinates $(t_{s}, {\bf x}_{s})$ until its lifetime $\tau_{s}$ is able to exclude this. Therefore, without contradicting ideas of a new Dirac equation (\ref{24}), the operators $\partial_{\mu}^{s}$ and $\partial_{\tau}^{s}$ can individually act on the wave function $\psi_{s}$ as well as on any of the existing types of fields. In our case, we are led to the following relations:
\begin{equation}
\partial_{\mu}^{s}\psi_{s}=\partial_{\mu}^{s}\psi_{s}(x_{s}), \, \, \, \, 
\partial_{\tau}^{s}\psi_{s}=\partial_{\tau}^{s}\psi_{s}(\tau_{s}),
\label{27}
\end{equation}
\begin{equation}
\partial_{\mu}^{s}\alpha_{s}=\partial_{\mu}^{s}\alpha_{s}(x_{s}), \, \, \, \, 
\partial_{\tau}^{s}\alpha_{s}=\partial_{\tau}^{s}\alpha_{s}(\tau_{s}).
\label{28}
\end{equation}

In the presence of the second type of a vector transformation (\ref{22}), the expected structure of the Lagrangian (\ref{25}) encounters appearance of an additional term and requires the restoration 
of its broken gauge invariance from the point of view of the origination of fermion mass at the 
level of the Dirac interaction. For this, we must at first introduce the photon Newton field $A_{\tau}^{s}(\tau_{s}),$ which at the availability of Eq. (\ref{22}) has a vector gauge 
transformation such as
\begin{equation}
A_{\tau}^{s'}=A_{\tau}^{s}+\frac{i}{m_{s}}\partial_{\tau}^{s}\alpha_{s},
\label{29}  
\end{equation} 
where $m_{s}$ characterize the Newton mirror interaction constants at the level of an electric 
mass of a vector nature.

To express the idea more clearly, it is desirable to present $\partial_{\tau}^{s}$ in the form
\begin{equation}
\partial_{\tau}^{s}=\partial_{\tau}^{s}-m_{s}A_{\tau}^{s}.
\label{30}  
\end{equation} 

Using Eqs. (\ref{11}), (\ref{30}) and having in mind Eqs. (\ref{27}) and (\ref{28}), one can find from Eq. (\ref{25}) that the Lagrangian $L^{D}$ invariant concerning the local gauge transformations 
(\ref{9}), (\ref{10}), (\ref{22}), and (\ref{29}) includes the following structural components 
of the C-invariant Dirac interaction:
$$L^{D}=L_{free}^{D}+L_{int}^{D}=$$
\begin{equation}
=i\overline{\psi}_{s}(\gamma^{\mu}\partial_{\mu}^{s}+\partial_{\tau}^{s})\psi_{s}-
ie_{s}j^{\mu}_{C}A_{\mu}^{s}-im_{s}j^{\tau}_{N}A_{\tau}^{s}.
\label{31}  
\end{equation} 

Coulomb and Newton parts of vector leptonic current, $j^{\mu}_{C}$ and $j^{\tau}_{N},$ respectively, 
are responsible for the structure of the united Dirac interactions as well as for their unified nature. They can therefore be defined by the following manner:
\begin{equation}
j^{\mu}_{C}=\overline{\psi}_{s}\gamma^{\mu}\psi_{s},
\label{32}  
\end{equation} 
\begin{equation}
j^{\tau}_{N}=\overline{\psi}_{s}\psi_{s}.
\label{33}  
\end{equation} 

Conservation of these types of currents follows from the fact that the matrices $\gamma^{\mu},$ $\partial_{\mu}^{s},$ and $\partial_{\tau}^{s}$ satisfy the commutativity conditions. As a consequence, any equation from
\begin{equation}
\partial_{\mu}^{s}j^{\mu}_{C}=0,
\label{34}  
\end{equation} 
\begin{equation}
\partial_{\tau}^{s}j^{\tau}_{N}=0
\label{35}  
\end{equation} 
testifies in favor of the availability of each of them. This property corresponds in nature 
to the coexistence law of the continuity equations.

We note also that in the Lagrangian (\ref{31}) the field $A_{\mu}^{s}$ and
\begin{equation}
A_{\tau}^{s}=\pmatrix{A_{\tau}\cr B_{\tau}}, \, \, \, \,
A_{\tau}=\pmatrix{A_{\tau}^{L}\cr A_{\tau}^{R}}, \, \, \, \,
B_{\tau}=\pmatrix{B_{\tau}^{L}\cr B_{\tau}^{R}}
\label{36}
\end{equation}
are predicted as the Coulomb and Newton components of the same vector photon field. 

Simultaneously, as is easy to see, vector electric mass $m_{s}$ and charge $e_{s}$ of any 
C-invariant Dirac particle appear in the Newton and Coulomb part dependence of 
its interaction with the same vector photon, respectively.

If now replace the mass square $m_{s}^{2}$ for $-\partial_{\tau}^{s}\partial^{\tau}_{s},$
Eq. (\ref{15}) with the aid of the latter is written as follows:
\begin{equation}
(\partial_{\mu}^{s}\partial^{\mu}_{s}-\partial_{\tau}^{s}\partial^{\tau}_{s})\varphi_{s}=0,
\label{37}
\end{equation}
in which it is definitely stated that the operators $\partial_{\mu}^{s}$ and $\partial_{\tau}^{s}$ 
can individually act not only on the fermion but also on the boson $\varphi_{s}$ field
\begin{equation}
\partial_{\mu}^{s}\varphi_{s}=\partial_{\mu}^{s}\varphi_{s}(x_{s}), \, \, \, \, 
\partial_{\tau}^{s}\varphi_{s}=\partial_{\tau}^{s}\varphi_{s}(\tau_{s}).
\label{38}
\end{equation}  

To the same Eq. (\ref{37}) one can also lead by another way starting from the free 
boson Lagrangian
\begin{equation}
L_{free}^{B}=\frac{1}{2}\varphi_{s}^{*}(\partial_{\mu}^{s}\partial^{\mu}_{s}-\partial_{\tau}^{s}\partial^{\tau}_{s})\varphi_{s}.
\label{39}  
\end{equation} 

Insertion of Eqs. (\ref{11}) and (\ref{30}) in Eq. (\ref{39}) together with Eqs. (\ref{28}) and (\ref{38}) convinces us here that the Lagrangian $L^{B}$ becomes invariant at the local gauge transformations (\ref{10}), (\ref{18}), (\ref{23}), (\ref{29}) and behaves as 
$$L^{B}=L_{free}^{B}+L_{int}^{B}=$$
$$=\frac{1}{2}\varphi_{s}^{*}(\partial_{\mu}^{s}\partial^{\mu}_{s}-\partial_{\tau}^{s}\partial^{\tau}_{s})\varphi_{s}+$$
$$+\frac{1}{2}[e_{s}(J_{\mu}^{C}A^{\mu}_{s}-J^{\mu}_{C}A_{\mu}^{s})-
e_{s}^{2}\varphi_{s}^{*}\varphi_{s}A_{\mu}^{s}A^{\mu}_{s}]-$$
\begin{equation}
-\frac{1}{2}[m_{s}(J_{\tau}^{N}A^{\tau}_{s}-J^{\tau}_{N}A_{\tau}^{s})-
m_{s}^{2}\varphi_{s}^{*}\varphi_{s}A_{\tau}^{s}A^{\tau}_{s}].
\label{40}  
\end{equation} 

We see that the vector fields $A_{\mu}^{s}(A^{\mu}_{s})$ and $A_{\tau}^{s}(A^{\tau}_{s})$ 
can, respectively, interact with the Coulomb $J^{\mu}_{C}(J_{\mu}^{C})$ and Newton 
$J^{\tau}_{N}(J_{\tau}^{N})$ currents 
\begin{equation}
J^{\mu}_{C}=\varphi_{s}^{*}\partial^{\mu}_{s}\varphi_{s}, \, \, \, \,
J_{\mu}^{C}=\varphi_{s}^{*}\partial_{\mu}^{s}\varphi_{s},
\label{41}  
\end{equation} 
\begin{equation}
J^{\tau}_{N}=\varphi_{s}^{*}\partial^{\tau}_{s}\varphi_{s}, \, \, \, \,
J_{\tau}^{N}=\varphi_{s}^{*}\partial_{\tau}^{s}\varphi_{s},
\label{42}  
\end{equation} 
forming the two components of the same vector boson current. Its conservation expresses herewith 
the idea of the coexistence law of the continuity equations
\begin{equation}
\partial_{\mu}^{s}J^{\mu}_{C}=0, \, \, \, \, \partial^{\mu}_{s}J_{\mu}^{C}=0,
\label{43}  
\end{equation} 
\begin{equation}
\partial_{\tau}^{s}J^{\tau}_{N}=0, \, \, \, \, \partial^{\tau}_{s}J_{\tau}^{N}=0.
\label{44}  
\end{equation} 

The origination of vector electric mass $m_{s}$ and charge $e_{s}$ of C-invariant types of 
bosons with a nonzero spin, as stated in Eq. (\ref{40}), must be accepted as a consequence of the 
mass-charge structure of their interaction with the field of emission of the same vector photon. 
In the Lagrangian (\ref{40}) appears in addition a square of each of electric mass $m_{s}^{2}$ 
and charge $e_{s}^{2}$ of the photon itself jointly with a square of the corresponding part from 
the Newton $A_{\tau}^{s}A^{\tau}_{s}$ and Coulomb $A_{\mu}^{s}A^{\mu}_{s}$ components of its field.

The absence of one of sizes of $m_{s}$ or $e_{s}$ would imply that both do not exist at all. Such 
a connection expressing the idea of the mass-charge duality [16] arises owing to an interaction
mass-charge nature. As a consequence, any element of each of Lagrangians (\ref{31}) and (\ref{40}) testifies in favor of the existence of all the remaining ones. 

Therefore, we must recognize that Eqs. (\ref{31}) and (\ref{40}) together with tensors
\begin{equation}
F_{\mu\lambda}^{C}=\partial_{\mu}A_{\lambda}^{C}-\partial_{\lambda}A_{\mu}^{C},
\label{45}
\end{equation}
\begin{equation}
F_{\tau\sigma}^{N}=\partial_{\tau}A_{\sigma}^{N}-\partial_{\sigma}A_{\tau}^{N}
\label{46}
\end{equation}
constitute the full Lagrangian of the unified field theory of C-even neutrinos and bosons 
with a nonzero spin at the level of the mass-charge structure of gauge invariance
$$L=i\overline{\psi}_{s}(\gamma^{\mu}\partial_{\mu}^{s}+\partial_{\tau}^{s})\psi_{s}+$$
$$+\frac{1}{2}\varphi_{s}^{*}(\partial_{\mu}^{s}\partial^{\mu}_{s}-\partial_{\tau}^{s}\partial^{\tau}_{s})\varphi_{s}-$$
$$-\frac{1}{4}F_{\mu\lambda}^{C}F^{\mu\lambda}_{C}+
\frac{1}{4}F_{\tau\sigma}^{N}F^{\tau\sigma}_{N}-
ie_{s}j^{\mu}_{C}A_{\mu}^{s}-im_{s}j^{\tau}_{N}A_{\tau}^{s}+$$
$$+\frac{1}{2}[e_{s}(J_{\mu}^{C}A^{\mu}_{s}-J^{\mu}_{C}A_{\mu}^{s})-
e_{s}^{2}\varphi_{s}^{*}\varphi_{s}A_{\mu}^{s}A^{\mu}_{s}]-$$
\begin{equation}
-\frac{1}{2}[m_{s}(J_{\tau}^{N}A^{\tau}_{s}-J^{\tau}_{N}A_{\tau}^{s})-
m_{s}^{2}\varphi_{s}^{*}\varphi_{s}A_{\tau}^{s}A^{\tau}_{s}].
\label{47}  
\end{equation} 

Thus, the Newton component of any type of an interaction gives a kind of inertial mass to all 
the interacting matter fields. This in turn indicates the existence of their corresponding charge 
as a consequence of the same interaction Coulomb part. In other words, each of fermions or bosons interacting according to the Lagrangian (\ref{47}) possesses the mass and charge.

However, it is known [17-19] that the standard electroweak model, by itself, does not involve 
gravity and thereby needs in substantiation of the dynamical origination of spontaneous gauge symmetry violation and of mass of matter fields from the point of view of their interaction 
with Higgs bosons, introduction [20] of which is compatible with exclusion in its construction 
of a fundamental role of gravity. At the same time, the nature itself is not in force to give an inertial mass to all the particles and fields regardless of gravity, which comes forward in it as 
a grand unification [21]. 

Therefore, it seems possible to use, for example, any of recent experiments [22,23] about a new particle as the first confirmation of the existence of leptonic and neutrino strings [3] and their birth in the Large Hadron Colliders and in the most diverse collisions as a unity [10] of mirror 
and gauge symmetry laws. A beautiful example is the left- and right-handed paraleptons [24] and 
their paraneutrinos [25] of vector electroweak currents
\begin{equation}
(l^{V}_{L}, \overline{l}^{V}_{R}), \, \, \, \,
(l^{V}_{R}, \overline{l}^{V}_{L}),
\label{48}
\end{equation}
\begin{equation}
(\nu_{lL}^{V}, {\bar \nu_{lR}}^{V}), \, \, \, \,
(\nu_{lR}^{V}, {\bar \nu_{lL}}^{V}).
\label{49}
\end{equation}

In the case of the photon currents, they correspond in nature to a formation of bosonic strings 
such as the following vector diphotons:
\begin{equation}
(\gamma^{V}_{L}, \overline{\gamma}^{V}_{R}), \, \, \, \,
(\gamma^{V}_{R}, \overline{\gamma}^{V}_{L}).
\label{50}
\end{equation}

There exists of course a range of bosons, which have the same spin, possess difference masses. 
In these differences, a crucial role of the dynamical distance [26] appears between the structural particles in paraparticles.

It states that each of Eqs. (\ref{48})-(\ref{50}) is conserved only if any action does not separate it by parts in the particle type dependence. Therefore, to understand the nature of a new particle [21,22] at a more fundamental dynamical level, one must use its leptonic decays [27] as a certain indication to the availability of an explicit possibility for the observation in Large Hadron Collider of a neutrino mode of its decay.

However, in spite of that the decay of each of Eqs. (\ref{48})-(\ref{50}) needs in special verification, the Lagrangian (\ref{47}) gives the right to consider a massive field as a consequence of the mass structure of gauge invariance.

\vspace{0.6cm}
\noindent
{\bf 4. Conclusion}
\vspace{0.3cm}

Here it is relevant to note that the Dirac equation (\ref{6}) one can define by inserting the Lagrangian (\ref{8}) in the Euler-Lagrange equation [28]. Of course, in a mirror presentation 
it has the form
\begin{equation}
\partial_{\mu}^{s}\left(\frac{\partial L_{free}^{D}}
{\partial(\partial_{\mu}^{s}\overline{\psi}_{s})}\right)=
\frac{\partial L_{free}^{D}}{\partial\overline{\psi}_{s}}.
\label{51}
\end{equation}

It is not surprising therefore that if Eq. (\ref{51}) is not in state to establish Eq. (\ref{24}) 
on account of Eqs. (\ref{25}) and (\ref{27}), we cannot exclude the existence of both Coulomb and Newton parts [15] in the same action. They constitute herewith the Euler-Lagrange equation at the level of the mass-charge nature of gauge symmetry
\begin{equation}
\partial_{\mu}^{s}\left(\frac{\partial L_{free}^{D}}
{\partial(\partial_{\mu}^{s}\overline{\psi}_{s}(x_{s}))}\right)+
\partial_{\tau}^{s}\left(\frac{\partial L_{free}^{D}}
{\partial(\partial_{\tau}^{s}\overline{\psi}_{s}(\tau_{s}))}\right)=
\frac{\partial L_{free}^{D}}{\partial\overline{\psi}_{s}(x_{s})}+
\frac{\partial L_{free}^{D}}{\partial\overline{\psi}_{s}(\tau_{s})}.
\label{52}
\end{equation}

Such a united state in turn implies that 
\begin{equation}
\partial_{\tau}^{s}\partial_{\mu}^{s}\psi_{s}(x_{s})=0, \, \, \, \, 
\partial_{\tau}^{s}\psi_{s}(\tau_{s})\partial_{\mu}^{s}\psi_{s}(x_{s})\neq 0,
\label{53}
\end{equation}
\begin{equation}
\partial_{\mu}^{s}\partial_{\tau}^{s}\psi_{s}(\tau_{s})=0, \, \, \, \,
\partial_{\mu}^{s}\psi_{s}(x_{s})\partial_{\tau}^{s}\psi_{s}(\tau_{s})\neq 0.
\label{54}
\end{equation}

For completeness we remark that the quantum energy and momentum operators from Eq. (\ref{5}) 
jointly with the size
\begin{equation}
E_{s}=\frac{{\bf p}_{s}^{2}}{2m_{s}}
\label{55}
\end{equation}
define in a mirror presentation the classical Schr\"odinger equation [29-32] as follows:
\begin{equation}
i\frac{\partial \psi_{s}}{\partial t_{s}}-
\frac{1}{2m_{s}\psi_{s}}\frac{\partial\psi_{s}}{\partial {\bf x}_{s}}
\frac{\partial\psi_{s}}{\partial {\bf x}_{s}}=0.
\label{56}
\end{equation}

Its unification with a particle quantum mass operator $m_{s}=-i\partial_{\tau}^{s}$
suggests one more naturally united equation
\begin{equation}
\partial_{t}^{s}\psi_{s}\partial_{\tau}^{s}\psi_{s}-
\frac{1}{2}\partial_{\bf x}^{s}\psi_{s}\partial_{\bf x}^{s}\psi_{s}=0
\label{57}
\end{equation}
in which each of operators $\partial_{t}^{s}$ and $\partial_{\bf x}^{s}$ similarly to all the operators $\partial_{\mu}^{s}$ and $\partial_{\tau}^{s}$ can individually act on the wave function
\begin{equation}
\partial_{t}^{s}\psi_{s}=\partial_{t}^{s}\psi_{s}(t_{s}), \, \, \, \,
\partial_{\bf x}^{s}\psi_{s}=\partial_{\bf x}^{s}\psi_{s}({\bf x}_{s}).
\label{58}
\end{equation}

Finally, insofar as the Newton components of weak and strong interactions and their inclusion 
in the Lagrangian uniting the electroweak and strong matter are concerned, all of them together 
with some aspects of the mass-charge structure of gauge invariance (not noted here) will be 
presented in the separate work.

\vspace{0.6cm}
\noindent
{\bf References}
\begin{enumerate}
\item
R.S. Sharafiddinov, Can. J. Phys. {\bf 93}, 1005 (2015); 1409.2397 [physics.gen-ph].
\item
R.S. Sharafiddinov, J. Phys. Nat. Sci. {\bf 4}, 1 (2013); physics/0702233.
\item
R.S. Sharafiddinov, Bull. Am. Phys. Soc. 57(16), KA.00069 (2012); 1004.0997 [hep-ph].
\item
R.S. Sharafiddinov, Fizika {\bf B 16}, 1 (2007); hep-ph/0512346.
\item
P.A.M. Dirac, Proc. Roy. Soc. Lond. {\bf A 117}, 610 (1928).
\item
P. Adamson et al., Phys. Rev. Lett. {\bf 110}, 251801 (2013). 
\item
M. Goldhaber, L. Grodzins, and A.W. Sunyar, Phys. Rev. {\bf 106}, 826 (1957). 
\item
M. Goldhaber, L. Grodzins, and A.W. Sunyar, Phys. Rev. {\bf 109}, 1015 (1958). 
\item
H. \"Uberall, Nuovo. Cim. {\bf 6}, 376 (1957).
\item
A. Giveon and E. Witten, Phys. Lett. {\bf B 332}, 44 (1994); hep-th/9404184.
\item
N.N. Bogoliubov and D.V. Shirkov, {\it Introduction to the Theory
of Quantized Fields} (Nauka, Moscow, 1984).
\item
O. Klein, Z. Phys. {\bf 37}, 895 (1926). 
\item 
W. Gordon, Z. Phys. {\bf 40}, 117 (1926-1927).
\item
W. Pauli and V. Weisskopf, Helv. Phys. Acta, {\bf 7}, 709 (1934). 
\item
R.S. Sharafiddinov, Bull. Am. Phys. Soc. 59(18), EC.00006 (2014); Spacetime Subst. {\bf 4}, 
235 (2003); hep-ph/0401230.
\item
R.S. Sharafiddinov, Bull. Am. Phys. Soc. 59(5), T1.00009 (2014); Spacetime Subst. {\bf 3}, 
47 (2002); physics/0305008.
\item
S.L. Glashow, Nucl. Phys. {\bf 22}, 579 (1961).
\item
A. Salam and J.C. Ward, Phys. Lett. {\bf 13}, 168 (1964).
\item
S. Weinberg, Phys. Rev. Lett. {\bf 19}, 1264 (1967).
\item
P.W. Higgs, Phys. Rev. Lett. {\bf 13}, 508 (1964).
\item
R.S. Sharafiddinov, Bull. Am. Phys. Soc. 60(4), E13.00008 (2015); hep-ph/0409254.
\item
The ATLAS Collaboration, Phys. Lett. {\bf B 716}, 1 (2012); 1207.7214 [hep-ex].
\item
The CMS Collaboration, Phys. Lett. {\bf B 716}, 30 (2012); 1207.7235 [hep-ex].
\item
R.S. Sharafiddinov, Eur. Phys. J. Plus, {\bf 126}, 40 (2011); 0802.3736 [physics.gen-ph].
\item
R.S. Sharafiddinov, Can. J. Phys. {\bf 92}, 1262 (2014); 0807.3805 [physics.gen-ph].
\item
V.G. Kadyshevsky, Nucl. Phys. {\bf B 141}, 477 (1978).
\item
The CMS Collaboration, Nat. Phys. {\bf 10}, 557 (2014). 
\item
V.S. Vladimirov, {\it Collection of Problems on Mathematical Physics Equations} 
(Nauka, Moscow, 1974).
\item
E. Schr\"odinger, Ann. der Phys. {\bf 384}, 361 (1926).
\item
E. Schr\"odinger, Ann. der Phys. {\bf 384}, 489 (1926).
\item
E. Schr\"odinger, Ann. der Phys. {\bf 385}, 437 (1926).
\item
E. Schr\"odinger, Ann. der Phys. {\bf 386}, 109 (1926).
\end{enumerate}
\end{document}